\pdfoutput=1

\documentclass[10pt,journal,compsoc]{IEEEtran}
\ifCLASSOPTIONcompsoc
  \usepackage[nocompress]{cite}
\else
  \usepackage{cite}
\fi
\ifCLASSINFOpdf
\else
\fi
\usepackage{amsmath}
\usepackage{array}
\usepackage{graphicx}
\usepackage{tabularx}
\usepackage{multirow}
\usepackage{xcolor}

\begin{document}
%
\title{A Data-driven Approach for\\ Furniture and Indoor Scene Colorization}
%
%
%
%

\author{
        {Jie~Zhu,
        Yanwen~Guo,
        and~Han~Ma}
        \thanks{Jie~Zhu, Yanwen~Guo,and Han~Ma are with the National Key Lab for Novel Software Technology, Nanjing University, Nanjing, China.
    E-mail:
        \protect{magickuang@126.com},
        \protect{ywguo@nju.edu.cn},
        \protect{1069595307@qq.com}
        }

}

\IEEEtitleabstractindextext{%
\begin{abstract}
We present a data-driven approach that colorizes 3D furniture models and indoor scenes by leveraging indoor images on the internet.
Our approach is able to colorize the furniture automatically according to an example image. The core is to learn image-guided mesh segmentation to segment the model into different parts according to the image object. Given an indoor scene, the system supports colorization-by-example, and has the ability to recommend the colorization scheme that is consistent with a user-desired color theme. The latter is realized by formulating the problem as a Markov random field model that imposes user input as an additional constraint. We contribute to the community a hierarchically organized image-model database with correspondences between each image and the corresponding model at the part-level. Our experiments and a user study show that our system produces perceptually convincing results comparable to those generated by interior designers.
\end{abstract}

\begin{IEEEkeywords}
Colorization, Interior Design, Data-driven Approach, Mesh Segmentation.
\end{IEEEkeywords}}

\maketitle

\IEEEdisplaynontitleabstractindextext

%
\IEEEpeerreviewmaketitle

\begin{figure*}[t]
    \centering
   \includegraphics[width=1.0\linewidth]{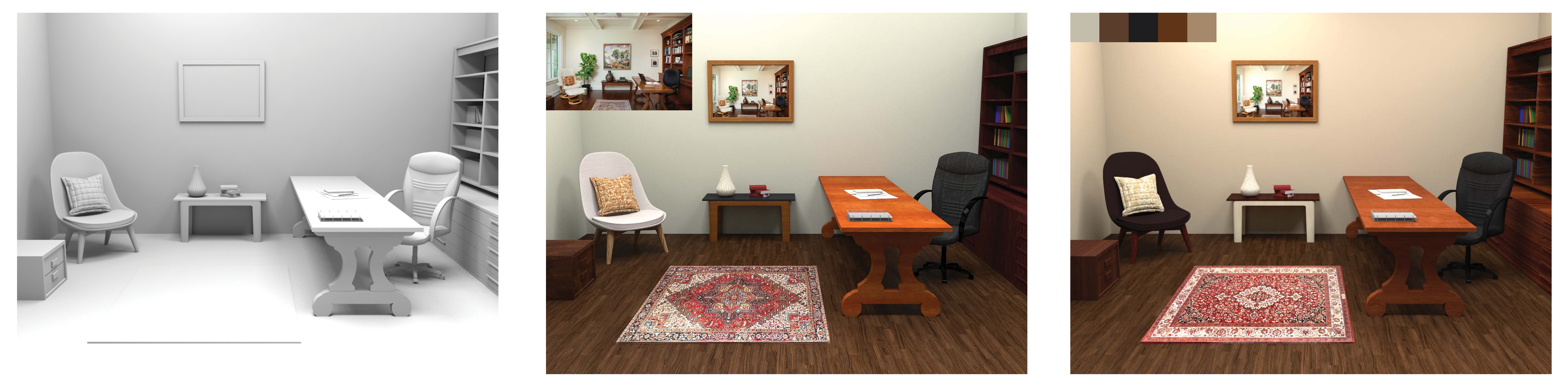}
   \caption{Left: A 3D indoor scene. Middle: The scene automatically colorized by our approach according to an example indoor image (inset).
   Right: Our recommended colorization scheme following a user-specified color theme (inset).}
   \label{TOPImage}
\end{figure*}

\IEEEraisesectionheading{\section{Introduction}\label{sec:introduction}}

%
%
%
%

\IEEEPARstart{``H}{ow} to choose the living room furniture, and how about this sofa?" Imagine you are visiting a furniture store or exploring the furnishing store online at Amazon. ``I have a rough idea of furniture arrangement but have no idea which one should I choose, the brown leather sofa or that one with the white and blue cover? So that it matches with the pink wallpaper and a warm and comfortable atmosphere could be created." Such questions may arise when you are ready to move into a new house and need to decorate it. Likewise, interior designers and game developers need to determine furniture colors after a new scene is built and the furniture layout is given.

Whereas in recent years progressive improvement has appeared demonstrating the capacity to help users create effective furniture layout~\cite{yu2011make,merrell2011interactive,fisher2012example}, automatic colorization of furniture objects and further the whole indoor scene has not yet received the considerable attention that it deserves.

Professional designers, having at least a few years of experience in interior design, usually rely on experience and intuition to choose colors for furniture and the whole scene. The aim is to ensure that color combinations are aesthetically appealing and visually pleasant, and the decoration is comfortable. For novices, this task is, however, challenging and usually requires tedious and time-consuming user intervention, especially for complicated scenes. Our goal is to develop algorithms capable of first, automatically colorizing furniture objects, and second, recommending to users visually pleasing colorization schemes for indoor scenes. This technique would be useful in not only interior design, but also other graphics applications requiring fully automatic scene modeling with a high degree of realism.

It should be noted that though scientifically accurate materials are essential for photorealistic rendering, the fact is that much of what people perceive and feel about the scene is experienced through colors and their combination. On the other hand, a specific furniture object only has a limited type of materials, for example, sofa is usually made of fabric or leather, and the dining table or a TV stand is often made by wood. Nevertheless, it could be painted with a broad range of colors. In this sense, it is easy to determine physical parameters for materials of different object categories~\cite{bell2013opensurfaces}, but determining colors can be a painful experience, especially when a lot of choices are available.

The Internet provides a large quantity of works by professional interior designers. This inspires us to colorize furniture by following the colorization schemes of these works. A major difficulty, however, lies in the fact that furniture models are often unstructured. As shown in Figure \ref{Sofatop}, the furniture model is composed of a set of topologically independent components, each of which is not necessarily to be semantically meaningful. Even though semantic segmentation applies to the model, the object is often not colorized completely according to different functional parts. In practice, the user needs to elaborately pick out those components with the same material to assign them the same texture, and this is a tedious and daunting task for those models with a lot of components.

In this paper, we first automate this process by building an image-model database and learning the image-guided furniture segmentation. The database is organized hierarchically. For each object in the training images, we exhaustively label its parts with different colors and correspondingly label the models in the same category.
For every object category, a classifier is learned for each different segmentation. Once learned, it can be applied to any new furniture models
to automatically label them, yielding the segmentation that facilitates automatic colorization.

Our method provides two ways to colorize 3D indoor scenes as shown in Figure~\ref{TOPImage}. For an indoor scene, the first way is to create the colorization scheme that visually resembles an input indoor image. This could be conveniently accomplished since our approach is object-based, and we can force each furniture model to closely follow colors/textures of the corresponding object in the image. The second way is to recommend the colorization scheme that is consistent with a user-desired color theme. In order to this, for each object category, we build Gaussian mixture models (GMMs) to model distribution of color schemes in the training images. We also build pair-wise GMMs to characterize the joint distribution of colorization schemes of any two different object categories. Given a new indoor scene,  a Markov Random Field (MRF) model is employed to characterize the probability that each object in the scene is assigned a specific colorization scheme. This MRF model also incorporates the constraint ensuring that colorization of the whole scene resembles the user-input color theme. The Markov chain Monte Carlo sampling is used for optimization, to suggest an optimized colorization scheme.

In summary, our system supports colorization of 1) individual furniture objects, 2) entire indoor scenes, and 3) 3D scenes by transferring color configurations from interest images.
Our contributions are in three-fold. First, we develop an automatic algorithm for colorizing furniture, allowing the input model to be unstructured and releasing the user from labor-intensive specification of the correspondence between model components and colored image regions. An image-guided furniture segmentation method is specifically tailored for this. Second, we present two ways: colorization-by-example, and a data-driven framework based on MRF for colorizing indoor scenes. Third, we contribute to the community a well-structured image-model database with detailed annotations and image-model correspondences.

\begin{figure*}[htbp]
  \centering
  \hspace{1cm}
  \includegraphics[width=1.0\linewidth]{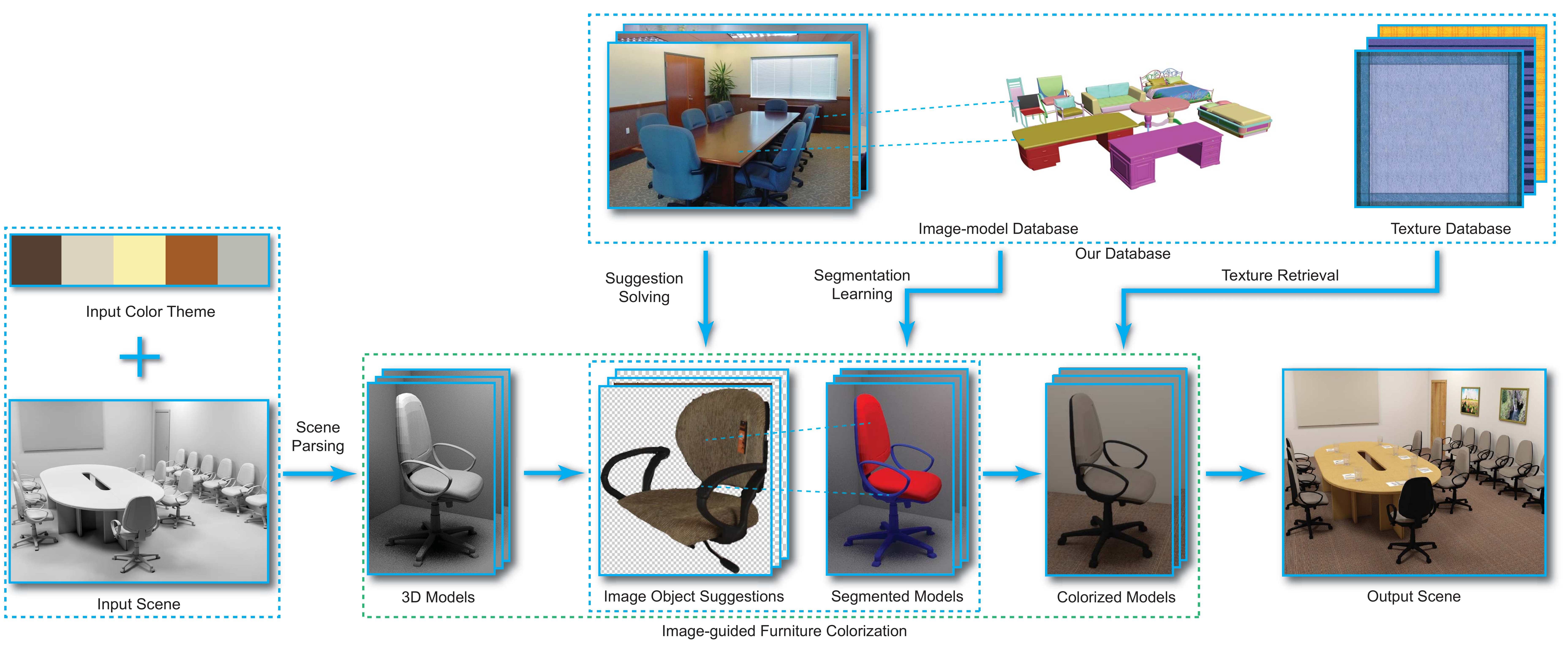}\\
  \caption{Pipeline of our 3D furniture and scene colorization framework.}\label{overview}
\end{figure*}

\section{Related Work}

\subsection{Image and Video Colorization}
Colorization, originally introduced to describe the computer-assisted process of colorizing those black-and-white movies, is now often used in Computer Graphics to describe the technique for adding colors to images and videos. Several advanced colorization techniques have been proposed in the past a few years.

Welsh et al. \cite{welsh2002transferring} described a semi-automatic technique for colorizing a grayscale image by transferring colors from a reference image.
Similarly, Irony et al. \cite{irony2005colorization} presented a novel method to colorize grayscale images by using a segmented example image.
In \cite{levin2004colorization},  the user indicates how each region should be colored by scribbling the desired color in the interior of the region. The colors are   automatically propagated to the remaining pixels in the image (sequence). Similar ideas have been adopted to colorize manga~\cite{qu2006manga} and natural images~\cite{luan2007natural}. Sykora et al. \cite{sykora2004unsupervised} introduced a novel colorization framework for processing old black-and-white cartoon videos. Chia et al.~\cite{chia2011semantic} proposed a colorization system that leverages the rich image content on the internet, and Lin et al.~\cite{lin2013probabilistic} colorized 2D patterns based on trained example patterns. Color removal, the inverse process of adding colors to grayscale images and videos has been studied in~\cite{gooch2005color2gray}.

Colorization is also related to the general recoloring problem which aims to adjust colors by using an input image as the reference~\cite{rasche2005re}, or a training set in a data-driven manner~\cite{cohen2006color,huang2014learning,yan2015automatic}.

\subsection{Mesh Colorization and Scene Decoration}
Inspired by image colorization, Leifman and Tal \cite{leifman2012mesh} proposed an algorithm for colorizing meshes. The user scribbles on the 3D model directly with  desired colors. The algorithm then completes colorization by propagating colors to the whole model. Such a method works well for an individual model, but is not
competent for indoor scenes with several furniture objects, in which scenario color compatibility among different objects is of vital importance. Furthermore, colorizing the furniture model by solely asking the users to indicate a few color samples without textures is insufficient to meet the stringent requirement of photorealistic rendering. By contrast, we realize furniture colorization under a data-driven framework.

The methods~\cite{jain2012material,nguyen20123d,chen2015decorator} on material suggestion for 3D models are most relevant to ours. Jain et al.~\cite{jain2012material} proposed to automatically assign 3D models materials. However, the requirement of a moderate number of high-quality 3D models as training data limits its application to indoor scenes, for which it is hard to collect a lot of training scenes with both high-quality geometry and material properties. The method proposed in~\cite{nguyen20123d} focuses on style transfer, and it cannot be used for applications that require a certain level of realism, for example interior design.
The method proposed in~\cite{wang2016unsupervised} focuses on extracting textures from an imaged object and transferring them to a 3D model. However, this method requires clean and independent imaged objects as input, so that it cannot be automatically applied to the imaged objects given an arbitrary indoor scene photograph.
The magic decorator system~\cite{chen2015decorator} automatically generates material suggestions for 3D indoor scenes which is casted as a combinatorial optimization considering both local material and global aesthetic rules defined.
Our approach has the following features, not possessed by this method. First, we are capable of colorizing individual furniture objects automatically. To automate this, we develop the image-guided furniture segmentation method with the support of our annotated image-model database. By contrast, given a new scene, this method does not provide the mechanism to add materials to each individual furniture object and to process the input scene with furniture models composed of unstructured components.
Second, we are able to transfer color configurations from interest images to indoor scenes, as shown in Figures~\ref{TOPImage} and ~\ref{Scenebyexample}. Last but more importantly, our approach is furniture-based, and this improves our flexibility as furniture is the basic element of a scene. As shown in Figure~\ref{refbedroom}, the user can interactively refine the initial result to change furniture colorization according to any interest images.

We are not aware of any other work that works directly for colorization of indoor objects and scenes. Interior designers and game developers normally rely on commercial tools, such as 3ds Max, and Maya, to accomplish this interactively. All these tools, however, require professional and high-level knowledge, and even for experts the colorization process is still a time-consuming task.

\subsection{Indoor Scene Modeling and Furniture Layout}
Our work is also inspired by the recent work on scene modeling and furniture layout. Motivated by the growing availability of 3D models along with model retrieval techniques, data-driven methods have been developed to reconstruct semantically indoor scenes from RGB-D images. These include \cite{nan2012search,kim2012acquiring,shao2012interactive,chen2014automatic}.

Besides, remarkable achievements have shown to help the users automatically synthesize indoor scenes populated by furniture objects~\cite{yu2011make,fisher2012example}, or interactively optimize an existing layout using interior design guidelines~\cite{merrell2011interactive}. The above methods focus on creating optimal furniture
arrangements, while are not capable of adding colors and materials to the scene.

\section{Overview}
Figure \ref{overview} provides an overview of our framework. Our input is a 3D indoor scene populated by furniture objects with known categories. We recommend appropriate color configurations for the furniture and the whole scene. Our approach works in a data-driven manner. We colorize each furniture model and the whole scene by learning from the large amount of indoor images taken by professional photographers or rendered by experienced interior designers that are available on the Internet. The framework consists of two key components: (1) colorization of each furniture object and (2) colorization of the whole scene.

\textbf{Colorization of an individual object}. Given a furniture model and a furniture image selected as the reference, we expect to render the model so that its colorization closely follows the image. A furniture model is usually composed by a group of topologically disconnected components, so the interior designer can pick some of them manually and apply a kind of material to them. Guided by the furniture image, our key is to pick out different groups of components automatically that should be consistently colored according to the corresponding image regions with different colors. For instance, the chair image in Figure \ref{overview} has two dominant colors, gray and black. How to automatically colorize the chair model according to the chair image remains a challenge.
This essentially can be viewed as a problem of model segmentation, which is guided by the imaged furniture.
We automate this by building an image-model database with exhaustively labeled mutual information between each furniture model and the corresponding furniture image at the \emph{part-level} and learning image-guided model segmentation. With the segmentation result, we could easily colorize the model with textures similar to the example image. This is done by searching from our material database the best matched textures whose dominant colors are consistent with colors of the imaged furniture. This step, illustrated in the green dashed box of Figure \ref{overview}, is described in detail in Section 4.

\textbf{Colorization of the whole scene}.
Based on the above method of furniture colorization, we could easily colorize the whole scene by making its colorization scheme perceptually resemble an example indoor image. Alternatively, we are able to recommend colorization schemes by incorporating users' preferences over colors.
For the whole scene, color combinations among different furniture objects are important to create an aesthetically pleasing and comfortable interior design.
We use a MRF model to characterize the probabilities of color configurations of all furniture and the probabilities of color combinations among different objects. The color suggestion for each furniture object is generated by sampling the density function of GMMs characterizing the distribution of color themes of the corresponding object category, and the state space of color combinations is modeled by GMMs as well. People often have a strong preference over particular colors, and they may choose their preferred themes to decorate the scene. To meet this requirement, we ask the user to provide a target color theme which is imposed as an additional constraint of our MRF model. The Markov chain Monte Carlo sampling algorithm is employed for optimization, yielding the colorization scheme that makes the scene look natural and perceptually close to the user-desired theme. We elaborate this step in Section 5.



\section{Image-guided Furniture Colorization}
 \begin{figure}[htbp]
  \centering
  \includegraphics[width=0.98\linewidth]{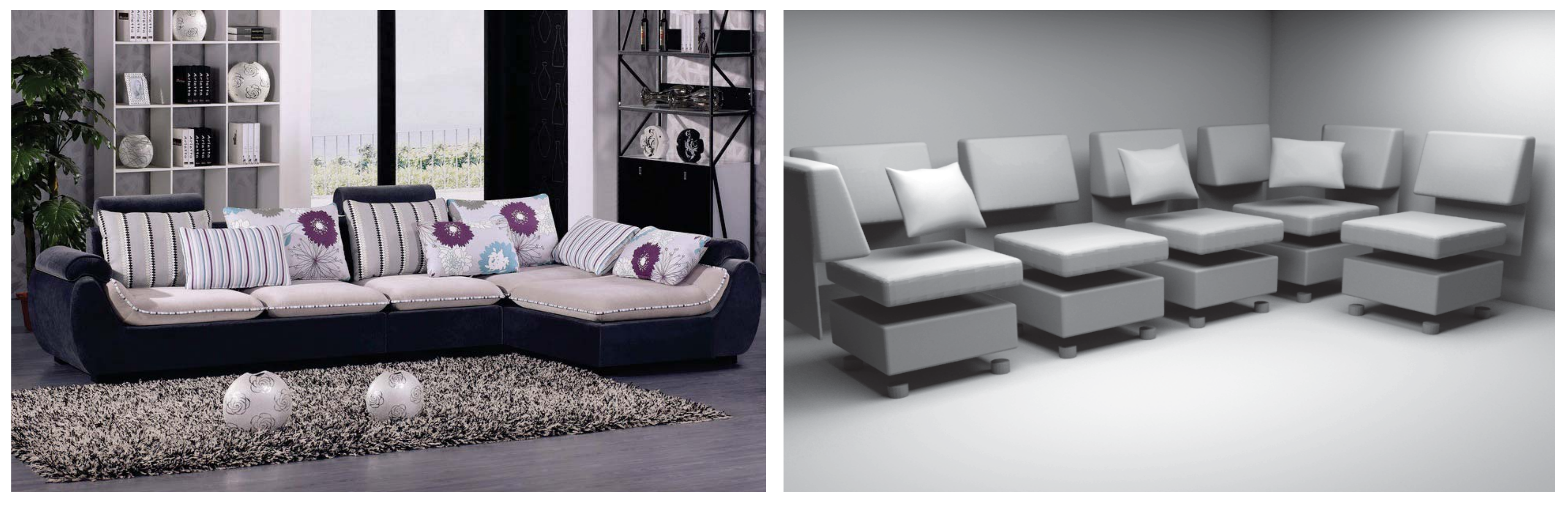}
  \caption{Left: A living room image with a modern sofa. Right: The sofa model has 40 topologically disconnected components.
 It is obvious that clear relations between these components and the sofa regions with different colors are not immediately available.
 Automatic colorization of the sofa model with the similar color theme as the image sofa is not trivial.}\label{Sofatop}
 \end{figure}

\subsection{Why Automatic Furniture Colorization is Not Trivial?}
Given a 3D furniture model, our goal is to generate an appropriate colorization scheme for this model by using an arbitrary image of furniture in the same category
so that the recommended colorization follows the image object. We call this image-guided furniture colorization. As shown in Figure~\ref{Sofatop}, a furniture model is usually composed of a group of topologically independent geometric components. Though the designers have already named each component in the process of geometric modeling, the names are, however, not always semantically meaningful. These pose challenges to users in that assigning different groups of components with different materials and colors using 3D modeling software such as 3ds Max could be boring and time-consuming.
This problem is an obstacle for an indoor scene colorization method. However, it remains unclear how this is handled by the Magic Decorator method~\cite{chen2015decorator}.

To tackle this problem, the key is to establish automatically the correspondences between those components and object regions with different colors. This could be viewed as a labeling problem. We resolve this problem by first building an image-model database. We then learn the image-guided mesh segmentation, and once learned, we could easily render the model whose colorization resembles the input image.

\subsection{The Image-model Database}
 \begin{figure}[h]
  \centering
  \includegraphics[width=1\linewidth]{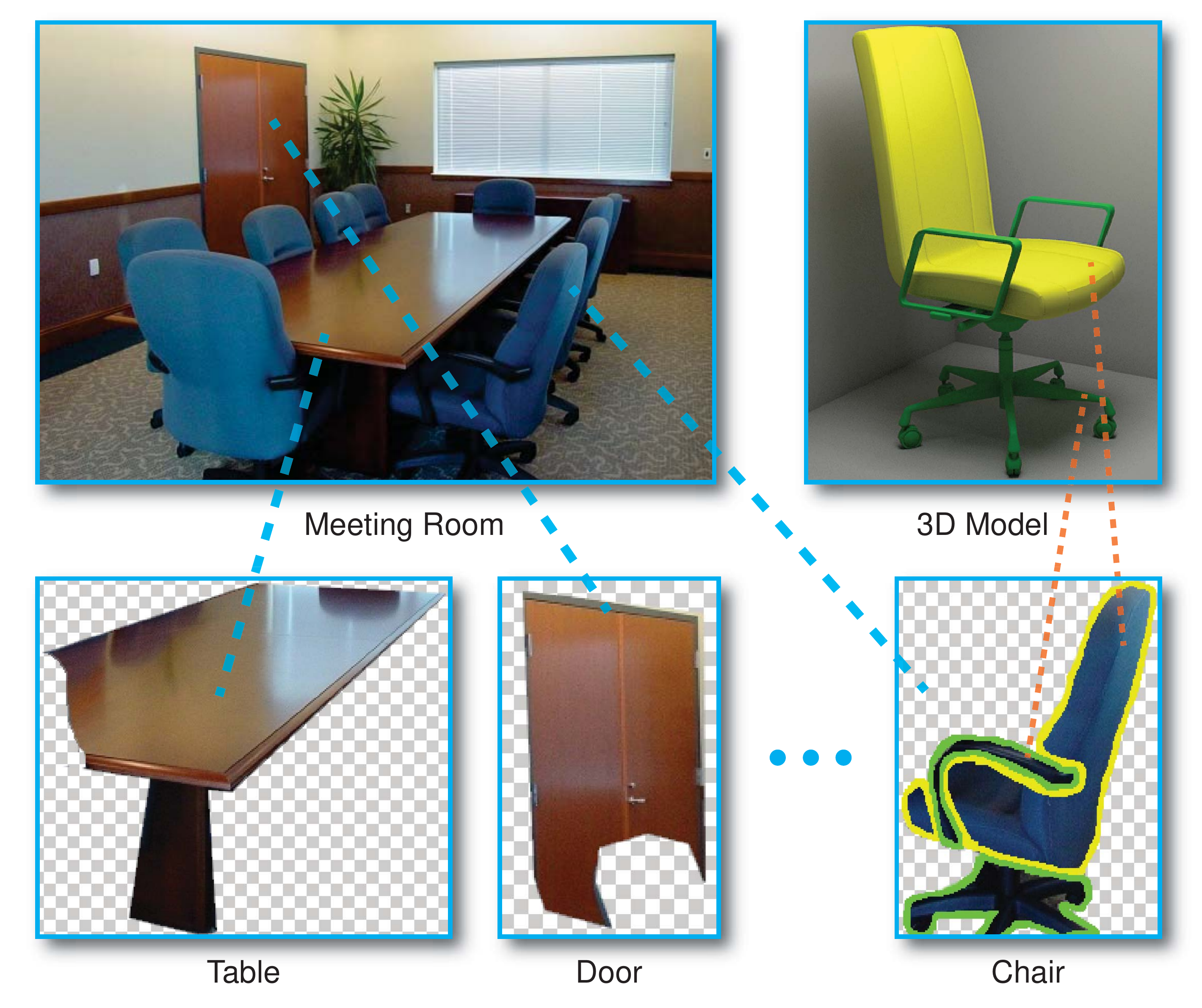}
  \caption{Each image is annotated with three-level information. For this image, the first level is ``meeting room", and the second is used to denote furniture objects it contains. For each object, the third level indicates its segmentation according to which those 3D models in the same category are processed.}\label{Labelhierarchy}
 \end{figure}
We have collected 1680 images of different scene categories from the Internet. As shown in Figure \ref{Labelhierarchy}, each image is annotated with three-level hierarchical information. The first level is the scene level in which each image is given a scene name, including the living room, bedroom, dinning room, office, and meeting room. The second is the object level. Each furniture region is annotated with a label according to its category, such as sofa, chair, coffee table, conference table, bed, floor, wall, and so forth. In the third level, the object region is divided into several parts according to different materials and colors. Each part is assigned a label and the associated material category. It should be noted that this label does not necessarily to be semantically meaningful. For instance, the arms and legs of an armchair should have the same label if they have the same material and colors. We have developed a prototype for image labeling. The prototype first over-segments the input image into regions without intensity boundaries~\cite{Felzenszwalb2004Efficient}. By doing so, we can easily annotate image regions by selecting those regions instead of pixels and only need to refine details when artifacts show up.
We show our image labeling tool in our supplemental material. It is user-friendly so the users can easily annotate their favorite images.

\begin{figure}[htbp]
  \centering
  \includegraphics[width=1.0\linewidth]{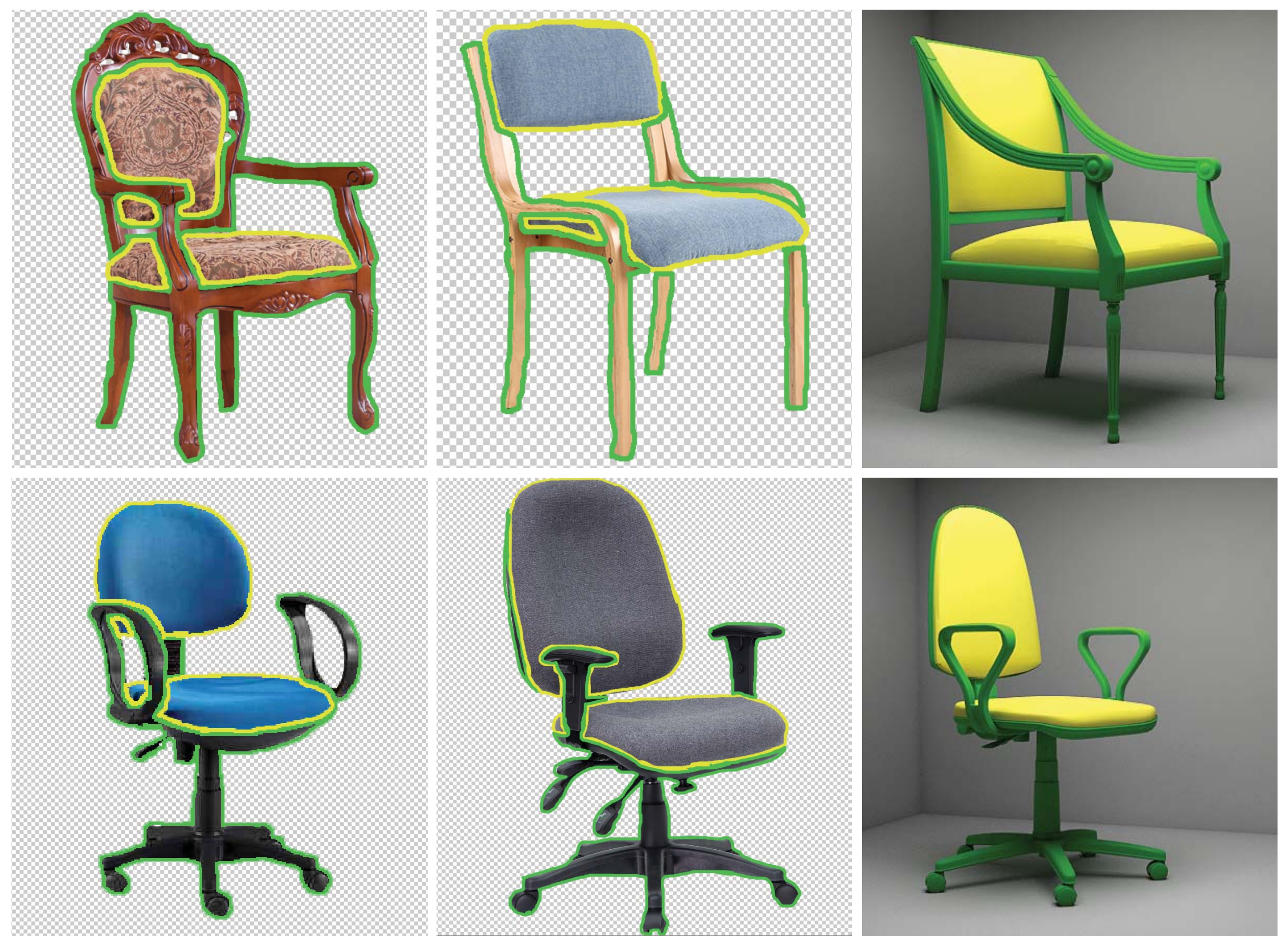}
  \caption{Up: The two chair images with similar segmentation according to which the chair model is labeled.
  Bottom: Another two chair images with similar segmentation according to which an office chair model is annotated.}\label{seg3}
\end{figure}

We also collected 200 3D furniture models
which have texture coordinates and component information,
and built hierarchical correspondences between each model and the image of furniture in the same category. These models cover all the furniture categories we use. For every category, each furniture model needs to be labeled once according to each different segmentation of the corresponding furniture  (see Figure~\ref{seg3}), as labeled in the third level in our image annotation.
3D model designers tend to define clear parts for their product to make them well structured and easy-to-edit. However the models downloaded from Internet repositories cannot always maintain the part information well. In these cases, we need to search for connected components to recover the part information. Since a meaningful part of a model could contain more than one disconnected component, the model could be over-segmented compared to the original parts of the model defined by the designer. But this can be easily handled by our mesh segmentation algorithm.

Texture map is crucial for photorealistic rendering. The sample regions of photographs do not suffice due to perspective distortion and compounded illumination. To tackle this problem, we further collected 2650 texture swatches to build a texture database. Each swatch is assigned with a material category, which is associated with the third level of image annotation.

\subsection{Image-guided Furniture Segmentation}
Given an arbitrary furniture model and a furniture image selected from our database as the reference, our goal is to colorize the model so that the colorization scheme closely follows the reference image. The furniture model is usually composed of disconnected components. By contrast, the imaged furniture may have several colored regions. The key is thus to establish the correspondences between these components and image regions with different colors so that different components are colored according to different furniture image regions. This essentially can be viewed as a classification problem on these components and further a model segmentation task with the aim to assign each component a label. It should be noted that the labels are not necessarily to be semantic meaningful. They are simply determined by furniture image regions with different colors.

Please note that though mesh segmentation has been extensively studied, previous methods mainly work on the primitive-level with the aim to assign a label to each individual triangular mesh \cite{lai2008fast,Shapira2008Consistent}. By contrast, we are facing a component-based mesh segmentation problem since 3D furniture is always composed of topologically disconnected components each of which further consists of triangular meshes. Furthermore, our segmentation is guided by the imaged furniture which is also different from the traditional part-based or semantic segmentation \cite{shamir2008survey,Kalogerakis:2010:labelMeshes}. To this end, we propose to learn furniture segmentation built upon both the mesh-level and component-level features.

Given a furniture model, we first compute a set of local features which include the curvature, PCA feature, Shape Diameter Function (SDF) feature~\cite{Shapira2008Consistent}, Average Geodesic Distance (AGD) feature~\cite{Hilaga2001Topology}, Shape Context feature~\cite{Belongie2002Shape}, and Spin Image feature~\cite{Johnson1999Using} for every triangular mesh. Each feature computed on the triangle is weighted by its face area, and further normalized by the sum of face areas of the mesh.
The details on these features can be found in \cite{Kalogerakis:2010:labelMeshes}.

We also compute a group of global geometric descriptors for each component. The Gaussian curvature, SDF feature, and AGD feature are used. For the Gaussian curvature, we first model the distribution of curvatures calculated on all vertices of a component as a histogram. The weighed average, median, variance, skewness, and kurtosis of curvatures, as well as the curvature histogram are used to form the descriptor of this component. We compute these for the other two features similarly.

Similar to~\cite{Kalogerakis:2010:labelMeshes}, we use JointBoost for our training. JointBoost is a boosting classifier that has the capability for classification with large numbers of features as input, and it performs feature selection automatically. Another advantage of JointBoost over other classifiers is that it has a fast sequential learning algorithm. Please refer to~\cite{torralba2007sharing} for the details of JointBoost.

To assign a label to each component, we resort to a voting strategy. For the input model, we train a local classifier $J_l$ for each triangle and a global classifier $J_g$ for each component $P$. The probability that $P$ has the label $l$ is then computed as,
\begin{equation}
\mathcal{P}(P,l) = \sum_{i}\frac{1}{Z}area(m_i)*J_l(m_i) + \lambda J_g(P)
\end{equation}
where $m_i$ represents the $i$-th triangle of the component, and $area(m_i)$ denotes its area. $Z$, as the normalization factor, is the sum of face areas of the component. Here $\lambda$ is a coefficient that balances the local and global classifiers. It was obtained as $0.4$ by optimizing performance against ground truth over the training models. We finally assign the label with the highest probability to the component.

\subsection{Furniture Colorization}
With the segmentation result, we can easily select for the model similar textures which make it look perceptually close to the reference object. We measure the color configuration of an image region with the color theme. Following the previous work on the study of color compatibility and enhancement~\cite{wang2010data,o2011color}, we represent the color theme as a color palette with a set of five colors. Widely used by artists and researchers, such a representation of color themes is visually intuitive and effective. Besides, its computation is fast. To extract the color palette for an image region, we cluster the pixels in the region by K-means. This yields a $K$-bins color histogram in which the five largest bins form the color palette. We empirically set the number of clusters $K$ to 50 and through experiments we found that the number around 50 does not influence the accuracy much.

For an object part, we retrieve the top $M$ most similar textures from the corresponding material category of our texture database by using the following distance metric,
\begin{equation}
D(C_O,C_T) = \sum_{k=1}^5 \min_{j} ||{C_O}_k-{C_T}_j||_2
\end{equation}
where $C_O$ and $C_T$ denote the color themes of an object part and the texture in our texture database, separately. ${C_O}_k$ is the $k$-th color entry of $C_O$, and ${C_T}_j$ is the $j$-th entry of $C_T$.
We assign the most similar texture sample to the corresponding segment of the furniture model and leave the rest of the $M$ texture samples to users in case they have some preferences.
As an alternative strategy here, we randomly choose one from the $M$ texture samples instead of the most similar one to achieve variability. In our implementation $M=10$.

Other important physical parameters, especially reflectance, can be easily obtained from the existing material database since only a very limited types of materials are commonly used for a specific kind of furniture.

\section{Scene Colorization}

\subsection{Colorization-by-example}
The input to our algorithm is a 3D indoor scene with known object categories and a user-preferred training image that contains the same object categories as the scene.
The scene can be automatically colorized by applying the above image-guided colorization to each of the furniture objects.

\subsection{A Data-driven Framework}
Now, we describe our data-driven framework for recommending colorization schemes for 3D indoor scenes in detail. Besides a 3D indoor scene with known object categories, the additional input is a user-specified target color theme, instead of the user-preferred image in section 5.1. The algorithm gives suggestions on color configuration for the scene such that for each model its colorization is natural and visually pleasing, and for the whole scene color combinations are compatible. In order to do so, we formulate this problem as a Markov Random Field (MRF) model. Each 3D furniture model in the scene is represented as a node in the graphic model and each image object in the same category is a potential state of the corresponding node. We form an edge between two furniture objects only when there exist images in our database that contain both objects.

The energy function is defined as,
\begin{equation}
E =  E_D + \beta E_S + \gamma E_C,
\end{equation}
where $E_D$, $E_S$, and $E_C$ represent the data term, smoothness term, and the constraint indicating user-preferred color theme over the scene, separately. $\beta$, and $\gamma$ are two weights balancing
the three terms. To emphasize the importance of the user-preferred color theme, we set them to $1$ and $10$ and kept them constant in all our experiments.

The data term $E_D$ evaluates whether the colorization scheme of each furniture model is commonly used in real life and visually pleasant. A color theme seldom seen will have a small probability. It is calculated as,
\begin{equation}
E_D = \sum_i log(G(C_{Mi}))
\end{equation}
where $C_{Mi}$ is the color theme for an individual furniture model. For each object category, we model the distribution of color themes with the Gaussian mixture models (GMMs) based on the objects in the training images. Recall that a color theme is represented by a color palette with a set of five colors, a color theme is thus a 15 dimensional vector based on the RGB color space. A single GMM is not sufficient to represent the distribution because the number of instances is too small compared with the feature space. To tackle this issue, we instead use 5 GMMs with 16 kernels in our implementation, each of which represents the distribution of one entry of the color theme and accumulate them when seeking the probability of a instance. We denote the probability of color themes of $C_{Mi}$ by $G(C_{Mi})$.

The second term $E_S$ can be regarded as a smoothness term. It is expressed as,
\begin{equation}
E_S = \sum_{ij} log(G(C_{Mi},C_{Mj})).
\end{equation}
It measures whether or not color combinations of two objects in the scene are frequently observed. We also use GMMs with 8 kernels in our implementation, to characterize the distribution of mutual relationships between color themes of two different objects that are present simultaneously in the same indoor images.

The third term $E_C$, which represents the user's constraint, ensures that the colorization scheme of the whole scene looks perceptually close to the user-input color theme $C_U$,
\begin{equation}
E_C = -\frac{1}{Z}\sum_i \sum_{k=1}^5 \min_{j} ||{C_U}_k-{C_{Mi}}_j||_2
\end{equation}
where $Z$ is the normalization factor, making the third term have the same magnitude as the first term.

Our goal is to maximize the above energy function. It is a high dimensional, non-convex combinatorial optimization problem which is difficult to solve analytically. The Markov Chain Monte Carlo (MCMC) methods are generally used for sampling from multi-dimensional distributions \cite{hastings1970monte,liu2008monte}, especially when the dimension is high as our problem. We thus use MCMC to obtain an optimal color configuration of the whole scene. MCMC is based on constructing a Markov chain that has the desired distribution as its equilibrium distribution. The optimization process works in an iterative manner. Given the current state of the variables, a candidate state
is generated for the next iteration. The optimization terminates when the energy measured above remains stable after a certain number of state sampling. We use a toolbox named UGM \cite{schmidt2010ugm} for MCMC sampling. Currently, the colorization scheme minimizing Equation 3 is taken as the solution.
Then the method described in Section 4 is used to segment the models in the scene and to retrieve textures according to the solved image objects.

\begin{figure}[t]
  \centering
  \includegraphics[width=1\linewidth]{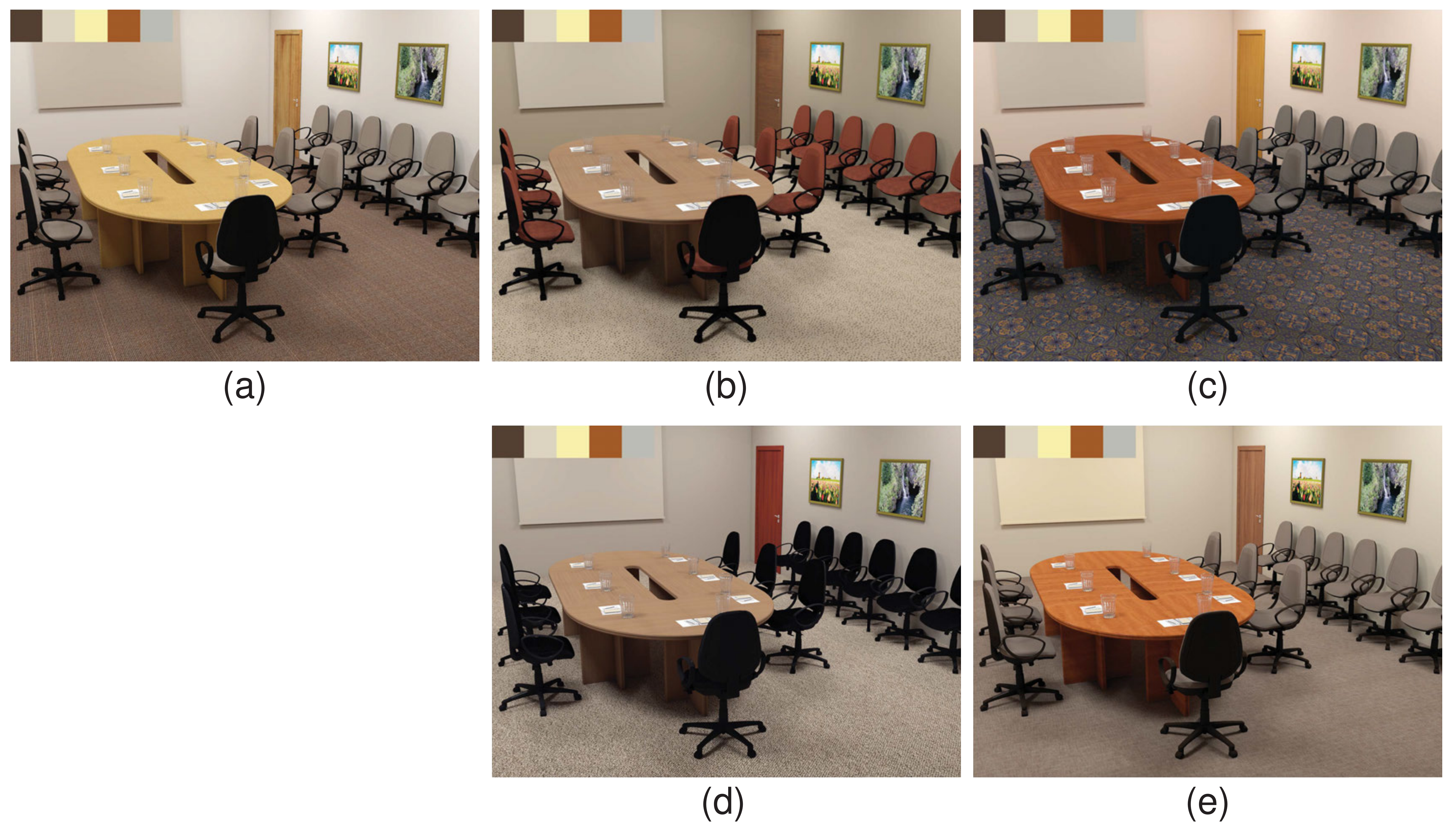}
  \caption{(a): The optimal solution by MCMC. (b) and (c): Other sampling results during the iteration of MCMC. (d): The solution ignoring user's constraint. (e): The solution considering user's constraint only.}\label{fig:MCMC}
\end{figure}

However, the user could select his/her favorite from multiple sampling results during the iteration, even though they are not the optimal solution to the energy function, as shown in Figure \ref{fig:MCMC} (b) and (c).

We also evaluate the user's constraint term $E_C$. As shown in Figure \ref{fig:MCMC}(d), without this term, the scene deviates from the target color theme much. In comparison, considering the user's constraint term only
makes the darker brown table obtrusive compared with its surroundings (Figure \ref{fig:MCMC}(e)), so the scene looks less harmonious than Figure \ref{fig:MCMC}(a).

\section{Experiments and Applications}

\subsection{Mesh Segmentation}

\begin{figure*}[htbp]
  \centering
  \includegraphics[width=1.0\linewidth]{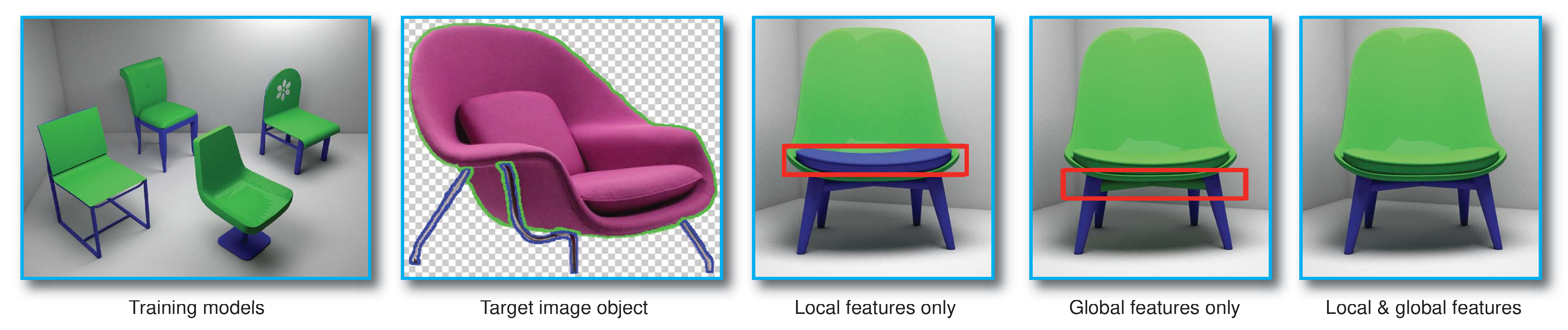}
  \caption{An example of image-guided mesh segmentation. The area in the red rectangle of the 3D model is incorrectly segmented if only the local classifier is used (3rd). Similarly, segmentation by the global classifier only is not right (4th). By combining the local and global classifiers, we obtain the segmentation that follows the target image.}\label{Meshseg}
\end{figure*}

Segmentation is a key component of our framework. We first verify its effectiveness.
Figure \ref{Meshseg} shows an example of our image-guided segmentation. Notice how the area in the red rectangle is incorrectly segmented if only the local or global classifier is used. A good segmentation is obtained by combining the local and global classifiers.

\begin{figure}[htbp]
  \centering
  \includegraphics[width=1\linewidth]{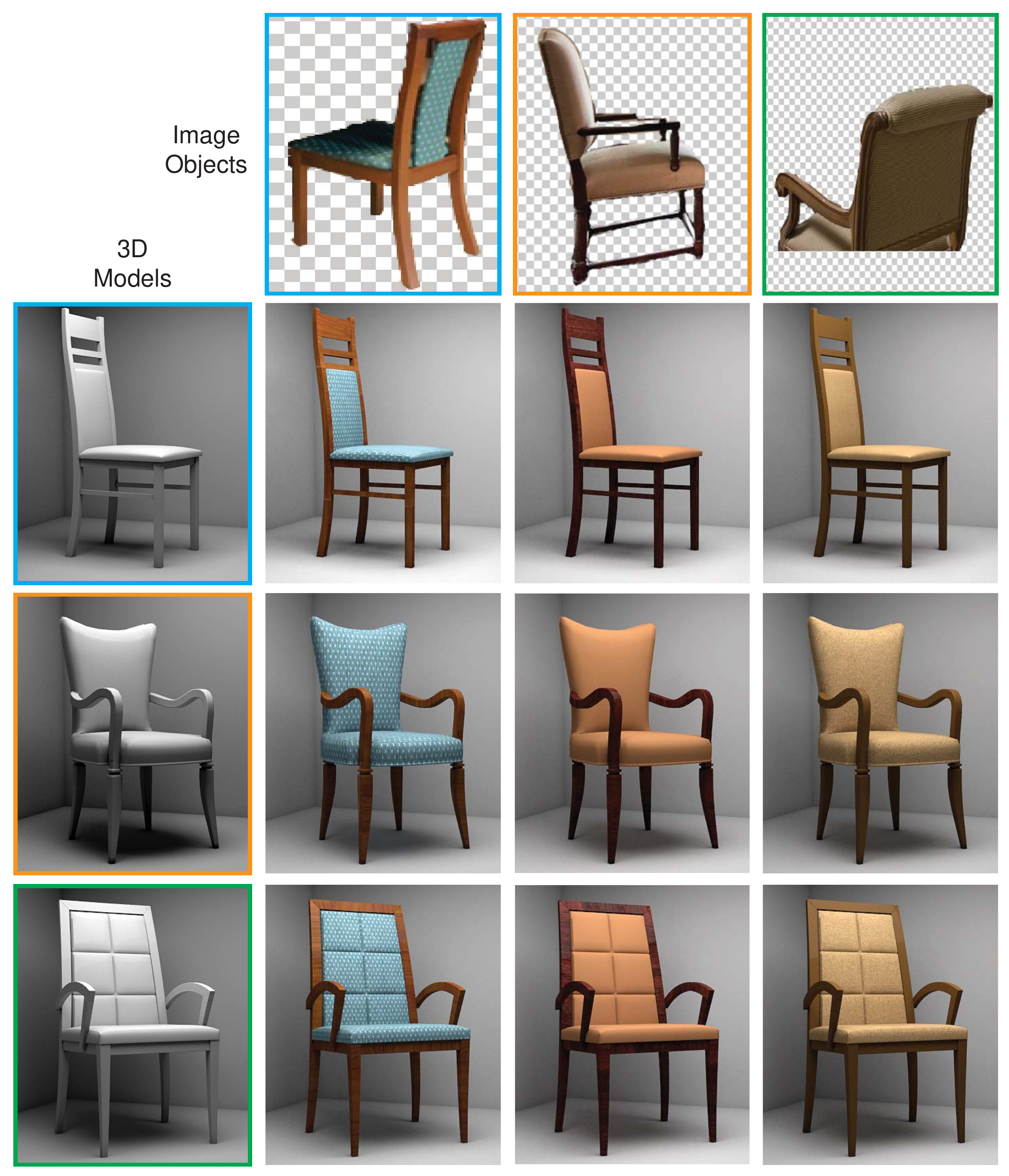}
  \caption{Results for three chair (armchair) models using three training images as the references.}\label{CFurniture1}
\end{figure}

\begin{figure}[htbp]
  \centering
  \includegraphics[width=1\linewidth]{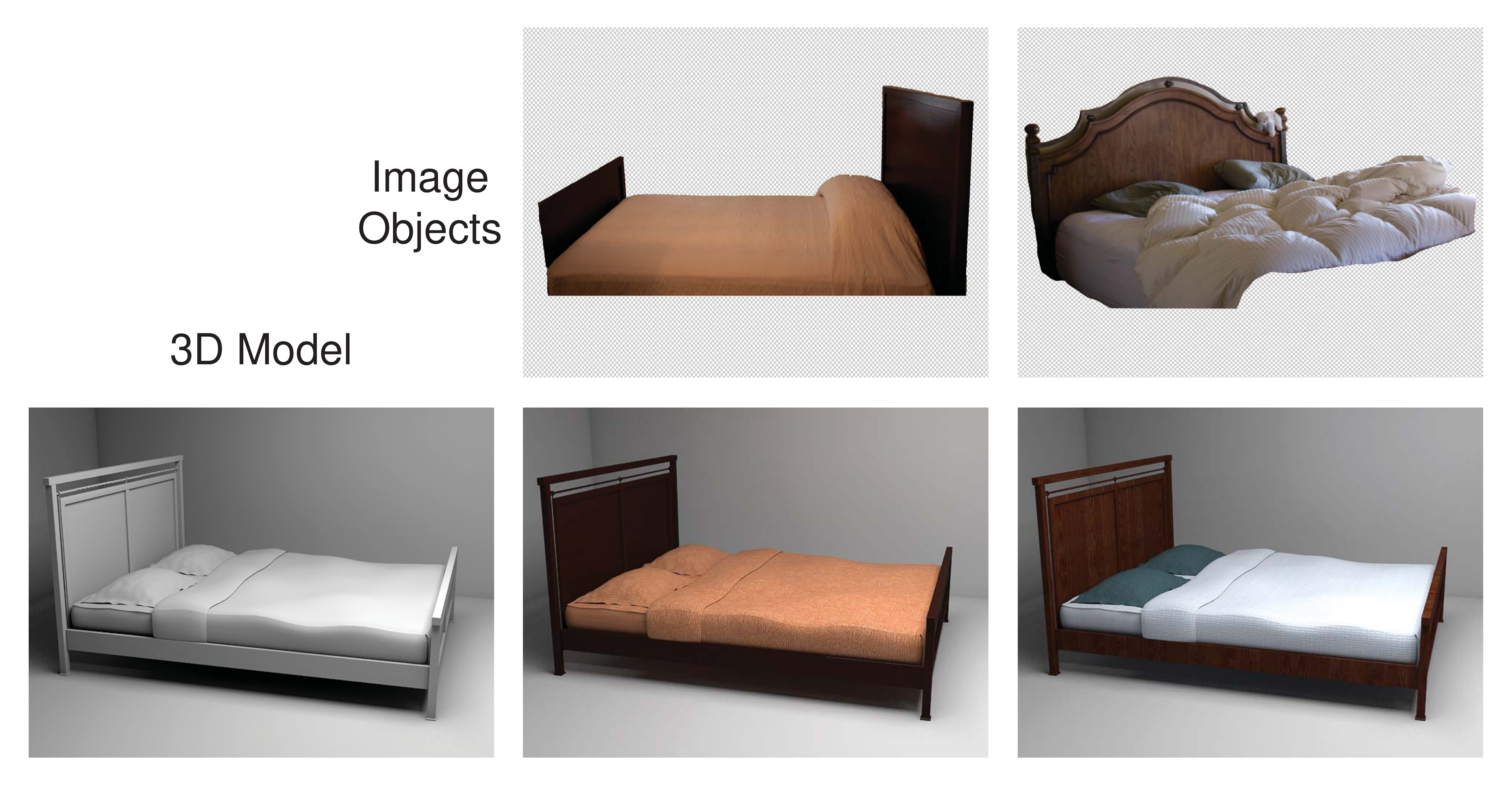}
  \caption{Results of a bed using two training images as the references.}\label{ColorizeBed}
\end{figure}

\begin{figure}[h]
  \centering
  \includegraphics[width=0.8\linewidth]{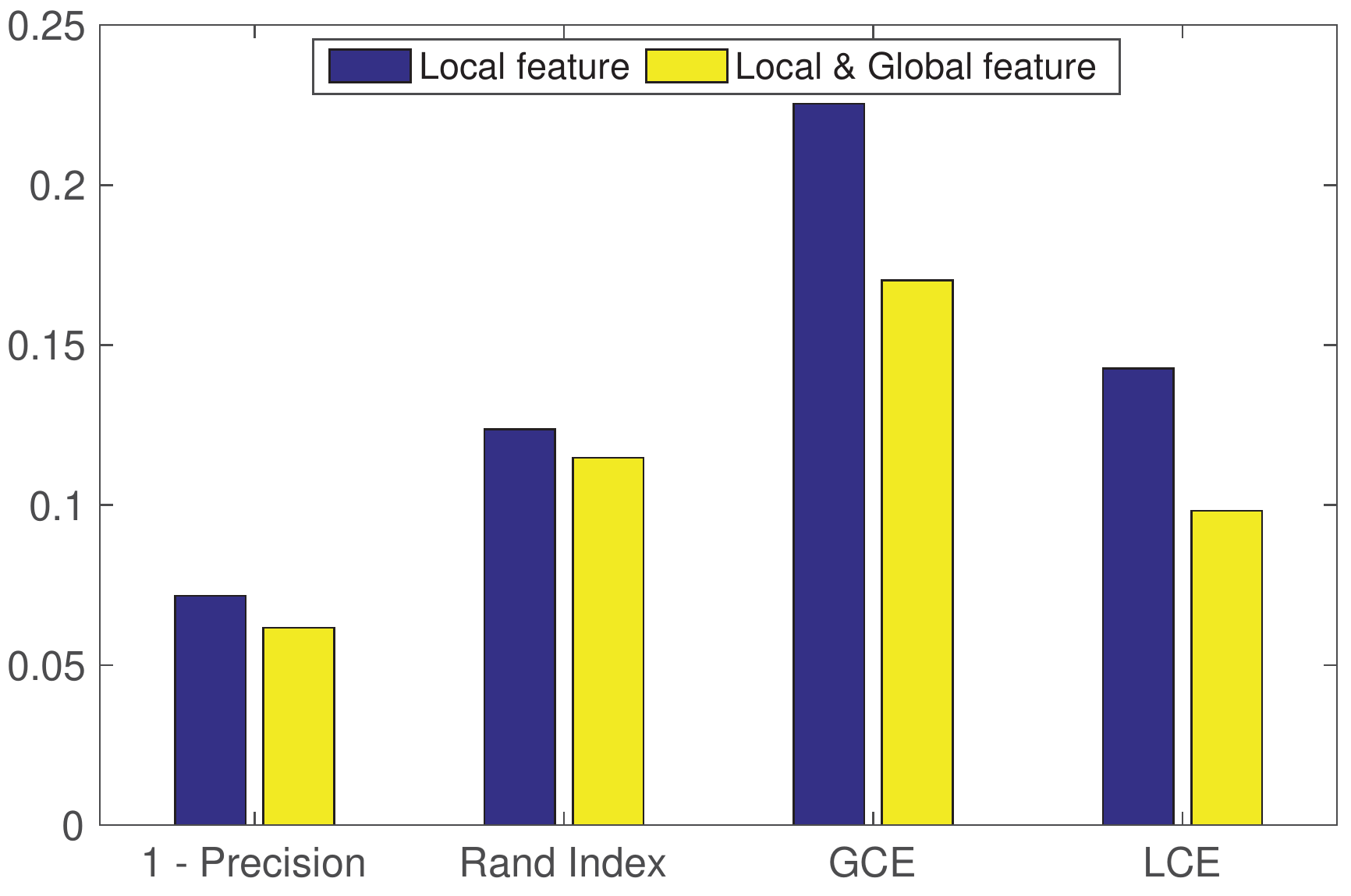}
  \caption{Evaluation of our image-guided mesh segmentation algorithm shows a significant improvement provided by combining the global geometric descriptors with the local descriptors.}\label{SegmentationEval}
 \end{figure}

We further qualitatively evaluate the segmentation performance.
In addition to the precision measure, Rand Index and Consistency Error have been used as metrics to evaluate segmentation performance \cite{Chen2009ABenchmark}. The former measures label consistency of a pair of faces between the segmentation result and ground truth segmentation, and the latter quantifies their hierarchical similarities which consists of two measures: Global Consistency Error (GCE) and Local Consistency Error (LCE). We modify them to fit our component-level segmentation. Totally 39 models selected from our model database are used as the training set and the rest are used for testing. We also report the 1-precision instead of precision for the purpose of being consistent with other metrics that represent dissimilarities rather than similarities. As shown in Figure \ref{SegmentationEval}, our method combining the global geometric descriptors with the local descriptors achieves a significant improvement over only using the local descriptors as the training features.
In the case of inaccurate mesh segmentation, the user has to manually correct it.

\begin{table*}[htbp]\footnotesize
\centering
\caption{Statistical data of scenes used in this paper.}
\begin{tabular}{|c|c|c|c|c|c|c|c|c|}
\hline
Scenes     & \begin{tabular}[c]{@{}c@{}}Office\\ (Fig. \ref{TOPImage})\end{tabular} & \begin{tabular}[c]{@{}c@{}}Meeting Rm.\\ (Fig. \ref{Scenebyexample}, row 2/3)\end{tabular} & \begin{tabular}[c]{@{}c@{}}Living Rm.\\ (Fig. \ref{Scenebyexample}, row 2/3)\end{tabular} & \begin{tabular}[c]{@{}c@{}}Dining Rm.\\ (Fig. \ref{Scenebyexample}, row 2/3)\end{tabular} & \begin{tabular}[c]{@{}c@{}}Dining Rm.\\ (Fig. \ref{RecResults})\end{tabular} & \begin{tabular}[c]{@{}c@{}}Living Rm.\\ (Fig. \ref{RecResults})\end{tabular} & \begin{tabular}[c]{@{}c@{}}Meeting Rm.\\ (Fig. \ref{RecResults})\end{tabular} & \begin{tabular}[c]{@{}c@{}}Bedroom\\ (Fig. \ref{refbedroom})\end{tabular} \\ \hline
Objects    & 10                                                       & 13/13                                                                  & 14/14                                                                 & 14/12                                                                   & 17                                                           & 11                                                           & 18                                                            & 7                                                           \\ \hline
Components & 43                                                       & 147/309                                                                & 91/53                                                                 & 162/93                                                                  & 97                                                           & 69                                                           & 258                                                           & 57                                                          \\ \hline
Materials  & 13                                                       & 8/8                                                                    & 16/15                                                                 & 11/10                                                                   & 11                                                           & 11                                                           & 8                                                             & 11                                                          \\ \hline
\end{tabular}
\label{my-label}
\end{table*}

\subsection{Furniture and Scene Colorization}

We test our method on typical furniture models and indoor scenes.
We would like to emphasize that all results on furniture and scene colorization shown in this section are generated automatically with our approach. The input to our system is the 3D furniture and scene models along with the furniture type, e.g., chair, sofa, or bed, and the scene type, e.g., living room, office, meeting room, and so on. Table~\ref{my-label} shows the statistical data of our scenes, including the number of objects, model components, and texture samples used for colorization.
Representative training images and models in our database are shown in the supplemental material.

\subsubsection{Furniture Colorization}

Figure~\ref{CFurniture1} shows our colorization results for three chair (armchair) models using three chair (armchair) images as the references. All armchairs in the training set is labeled as ``chair" in our database. In this example, we apply the same classifier to three models with different styles. Our approach produces colorization schemes that closely follow the three training images, even though segmentation results are not consistent across the three models. The nine colorized chairs look natural and visually pleasing.

Figure~\ref{ColorizeBed} shows the results for a bed model using two bed images as the references. Note that in the second column,
the bed model is segmented into two parts which correspond to the bed and bedclothes in the first reference image. By contrast in the third column,
it is segmented into three parts where the two pillows are segmented out by our image-guided segmentation with a different classifier.
It should be noted that in all our results, each bed model, the pillows and bedclothes on it always belong to a unified 3D model with a lot of disconnected components.
In our image-model database, ``pillow"(``bedclothes") belonging to ``bed" is not annotated as an independent object.
Our approach, however, is able to colorize the bed, pillows, and bedclothes with different colors automatically.

\subsubsection{Scene Colorization}

\begin{figure*}[htbp]
  \centering
  \includegraphics[width=0.95\linewidth]{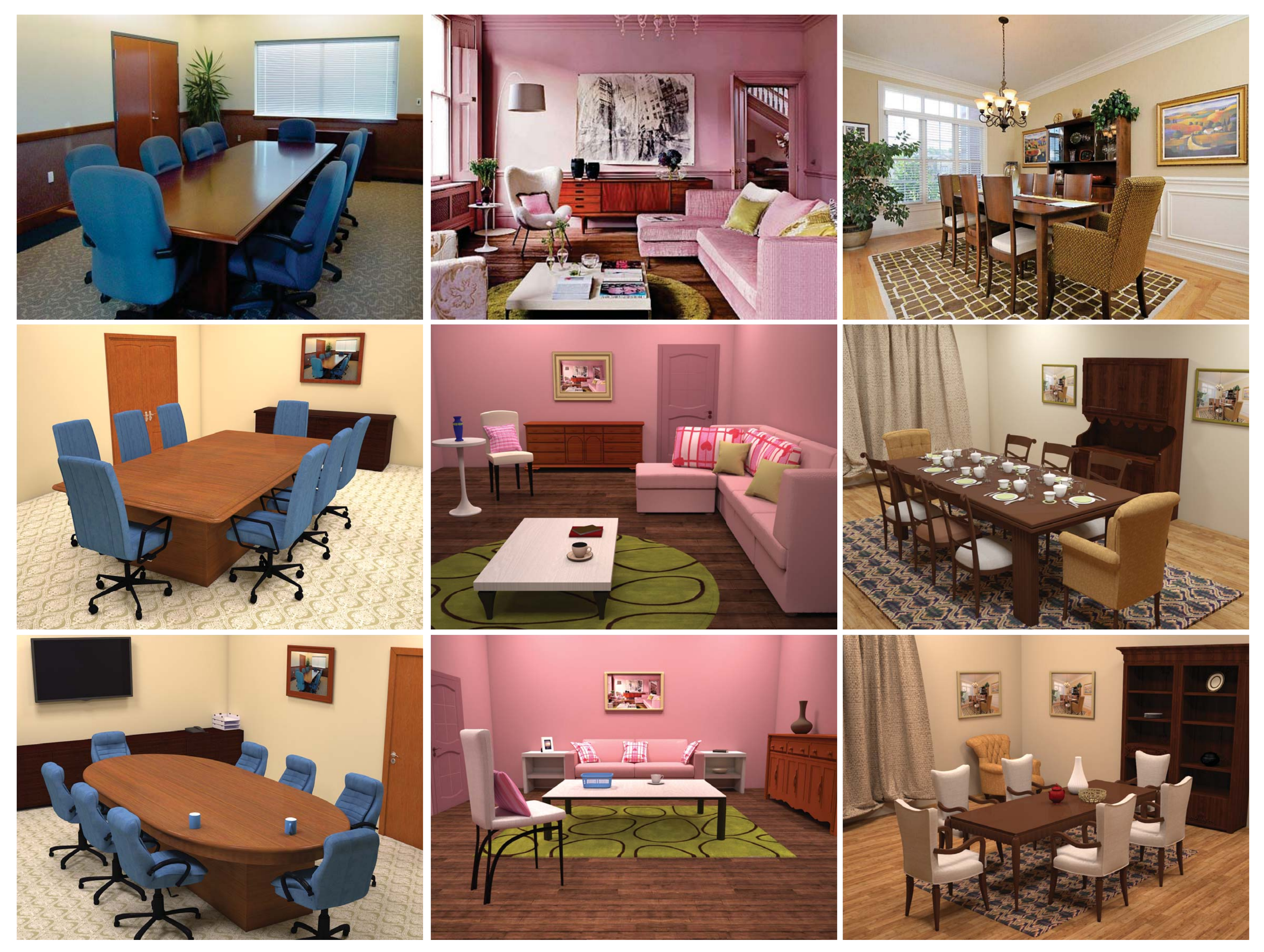}
  \caption{A distinctive feature of our approach is the ability to colorize 3D indoor scenes (2nd and 3rd rows) according to the reference images (1st row).}\label{Scenebyexample}
\end{figure*}

\begin{figure*}[htbp]
  \centering
  \hspace{1cm}
  \includegraphics[width=1\linewidth]{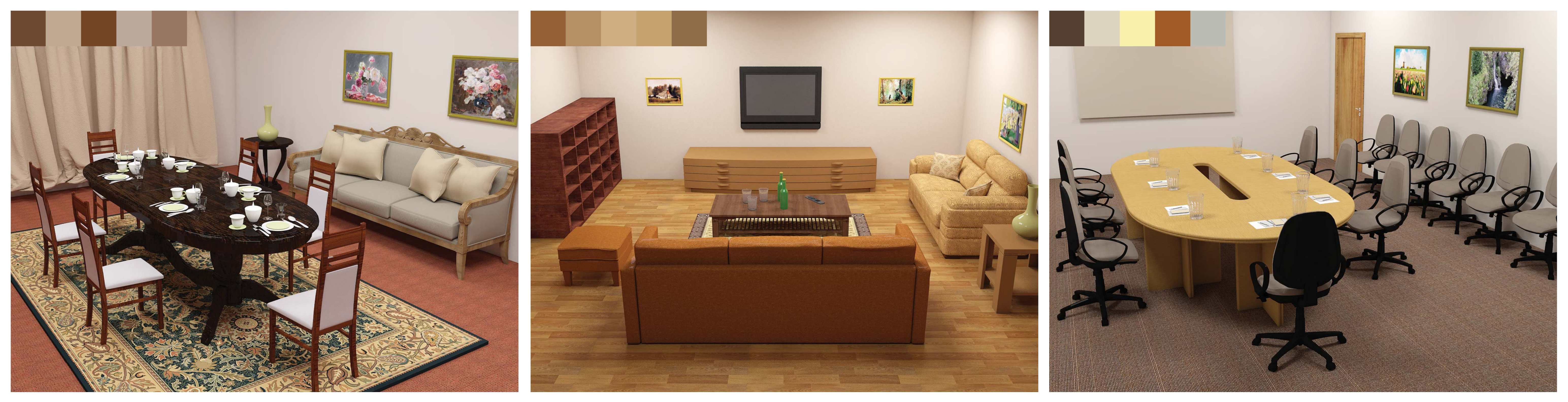}\\
  \caption{Our recommended colorization results with the user-input color themes as constrains (inset in each image).}\label{RecResults}
\end{figure*}

\begin{figure*}[htbp]
  \centering
  \hspace{1cm}
  \includegraphics[width=1\linewidth]{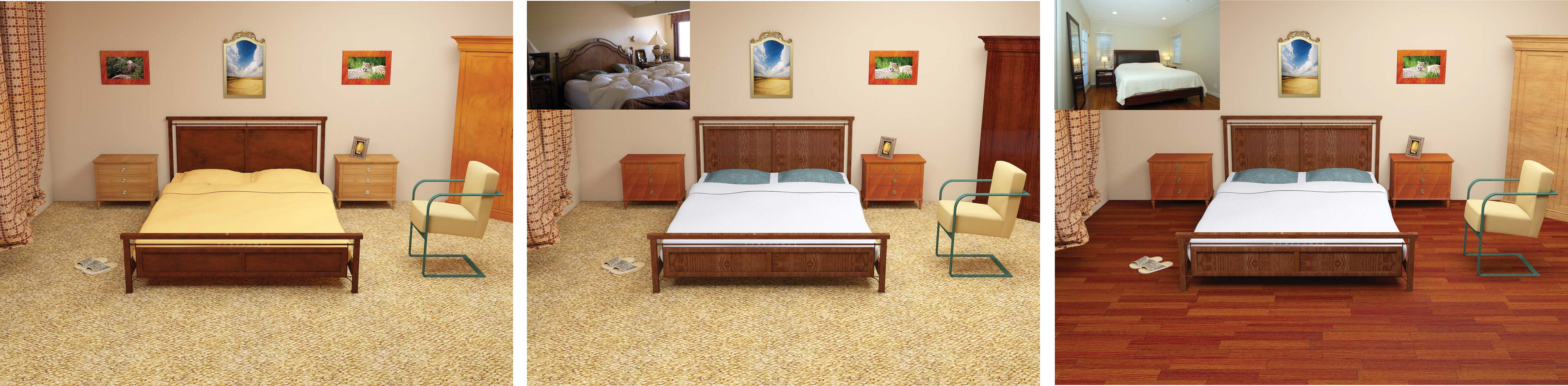}\\
  \caption{The user interactively refines the initial result of a bedroom (left). First, the user changes the bed color by using the bed of a bedroom image (the inset at the top left of the middle image) as the reference, and the scene is updated automatically (middle). The user further changes the initial carpet with wooden floor according to another bedroom image, leading to a new result (right).}\label{refbedroom}
\end{figure*}

The object-based property endows our approach the ability of colorization-by-example. That is, we can colorize the input scene by transferring color configurations from a user-provided reference - an example indoor image. We demonstrate this capacity with several results as shown in Figure~\ref{Scenebyexample}.
The first row shows the reference images. The second row shows the corresponding results where furniture arrangements resemble the arrangements in the corresponding input images. In the third row, we show the results for indoor scenes with different furniture models and arrangements.

Figure~\ref{RecResults} first shows our recommended colorization result for a dinning room given a user input color theme (inset).
The dining room is populated by a dining table, six chairs, a round table, a European style sofa and the throw pillows on it.
The carpet, curtain, and wall are automatically colorized as well since each of them has an independent label during image annotation, whereas the dinnerware, flower vase, and
two framed watercolor paintings are placed by the user.
Then we show in Figure~\ref{RecResults} the result for a living room which includes a TV stand, a cabinet, a coffee table, a 2-piece fabric sofa set (brown), a 2-seat fabric sofa with two throw pillows, and a side table. The wooden floor is colorized automatically as well. The last is a typical meeting room.

\textbf{Interactive Refinement.}
The user may wish to refine the initial colorization scheme with respect to his/her preference, for instance, changing the colors of a specific furniture object. This is conveniently supported by our framework since our approach is object-based.
Given a user-specified color theme as input, Figure \ref{refbedroom} first shows the result for a typical bedroom. In the next step, the user changes color configuration of the bed from dark red to brown wood, making it follow colorization scheme of the bed in a bedroom image. Our system then suggests new color configurations for most of the furniture models so that colorization scheme of the whole scene is compatible with the changed bed. The user further changes the carpet made of natural jute to the real wooden floor. To accommodate such a change, the wardrobe changes to a brighter color. By doing so, scene contrast is maintained, and  color combinations are harmonious and pleasing.

\subsubsection{Practicality}

For all our experiments shown above, the reference images are selected from the training set which are pre-segmented in the training stage. But if the user does not like the training images, a new example image is required. In such a case, the user needs to manually select the corresponding classifier, if the segmentation guided by this new image differs from all our pre-learned segmentations (i.e. classifiers trained on our dataset). Furthermore, we probably need to manually segment the new image and learn a new classifier.

\begin{figure*}[htbp]
  \centering
  \hspace{1cm}
  \includegraphics[width=1\linewidth]{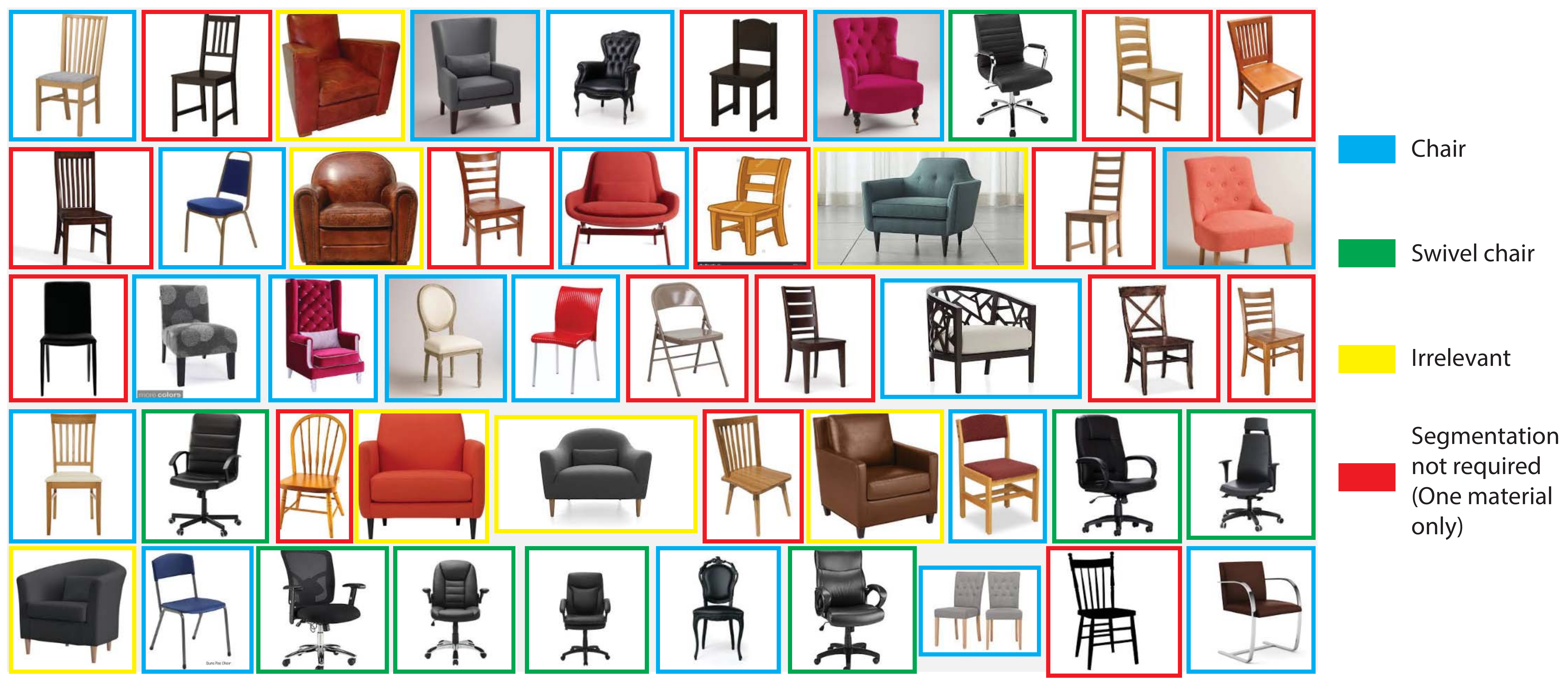}\\
  \caption{Images searched on Google Image with the keyword ``chair".}\label{RefChairSearch}
\end{figure*}

However, our dataset currently contains nearly $2000$ images which cover most different segmentations for different kinds of furniture. We searched images on Google Image by using different keywords on furniture to testify the practicality of our system. Figure~\ref{RefChairSearch} shows the one-screen searching results with the keyword ``chair". These searching results could be the potential user input of our system. Most of them are two types of chairs. Among them, ordinary chairs are marked with blue, and swivel chairs are marked by green. We examine these chair images one-by-one, and found that we already have the corresponding pre-trained segmentation classifiers in our database. The images marked by yellow are irrelevant to the keyword ``chair". They are actually sofa, and they correspond to ``sofa" in our dataset, not ``chair". The other images are marked by red, each of which only contains one material. Segmentation is not required when taking either of them as the example.

This shows for most cases, the users do not need to train new classifiers for segmentation, even though he/she prefers to use other images not included in out database as the example. More searching results with other keywords are shown in the supplemental material.

\begin{figure}[h]
  \centering
  \includegraphics[width=0.9\linewidth]{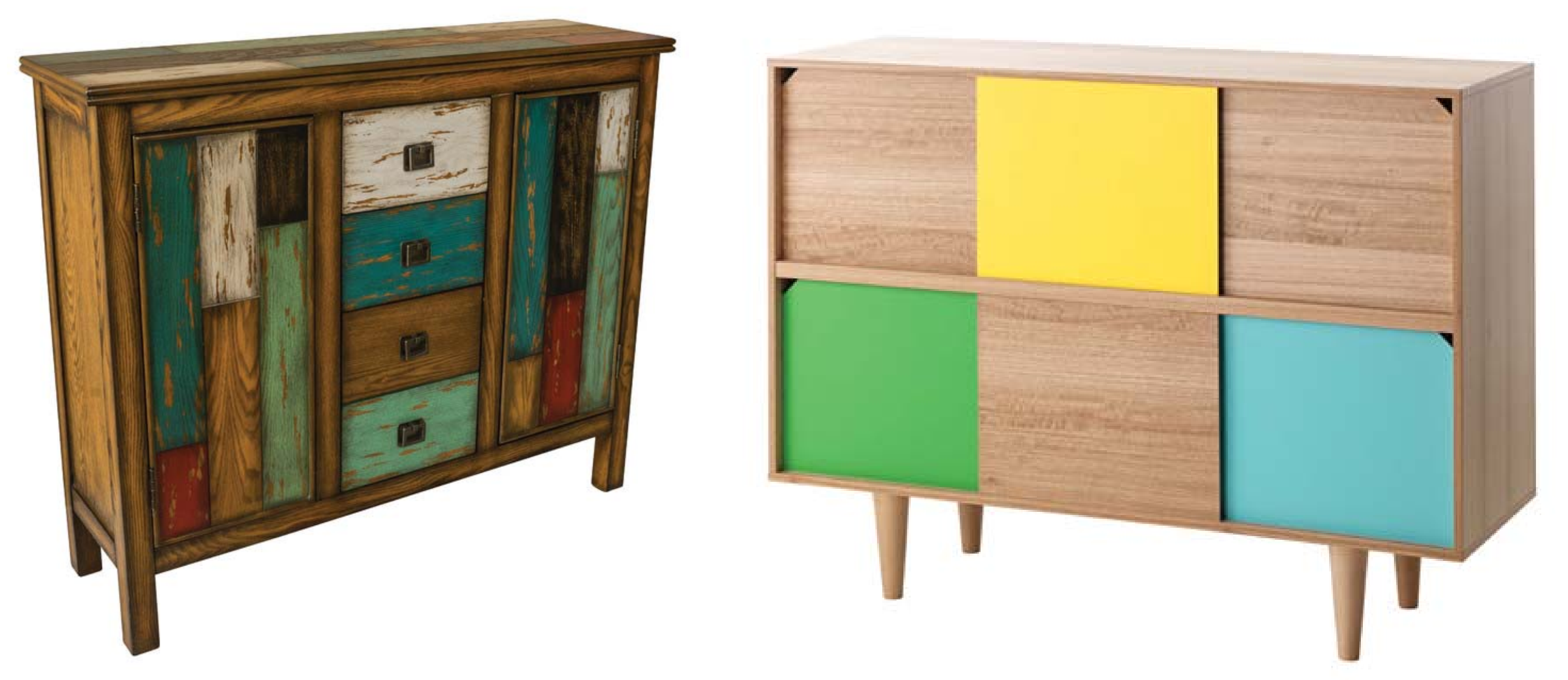}
  \caption{Examples of color patterns on the doors of the cabinets.}\label{Failcase}
 \end{figure}

\textbf{Limitations.}
We have to manually correct inaccurate segmentation, though seldom encountered. This is a limitation of our method.
However, as reported in Section 6.1, segmentation errors are seldom encountered during experiments.
Since our method is object-based, our segmentation method does not consider the color patterns among similar components of an image object. Figure \ref{Failcase} shows two examples with those patterns. In each of the two cabinets, the doors have similar shapes, but different colors.
Currently, our system is not capable of segmenting the mesh according to the colored components with similar geometry for this kind of objects. This is another limitation of our method.

An intrinsic image decomposition method could be helpful to our image database, for extracting lighting-free textures to be further used in rendering colorized scenes. However, such methods are not so robust that can be directly applied to various images in a large image database. On the other hand, intrinsic image decomposition is not essential to achieve good results in our experiments. So we did not incorporate it in our work, but we will further study it to improve our database.

\section{User Study}
We also devised a user study to measure how well users can differentiate between scene colorization results generated by our approach and those by interior designers.
For the study, we show each participant a sequence of rendered scenes. The images are always shown in groups and each group contains
three rendered results on exact the same 3D indoor scene by our approach and by two invited interior designers, separately.

\begin{figure*}[htbp]
  \centering
  \includegraphics[width=0.9\linewidth]{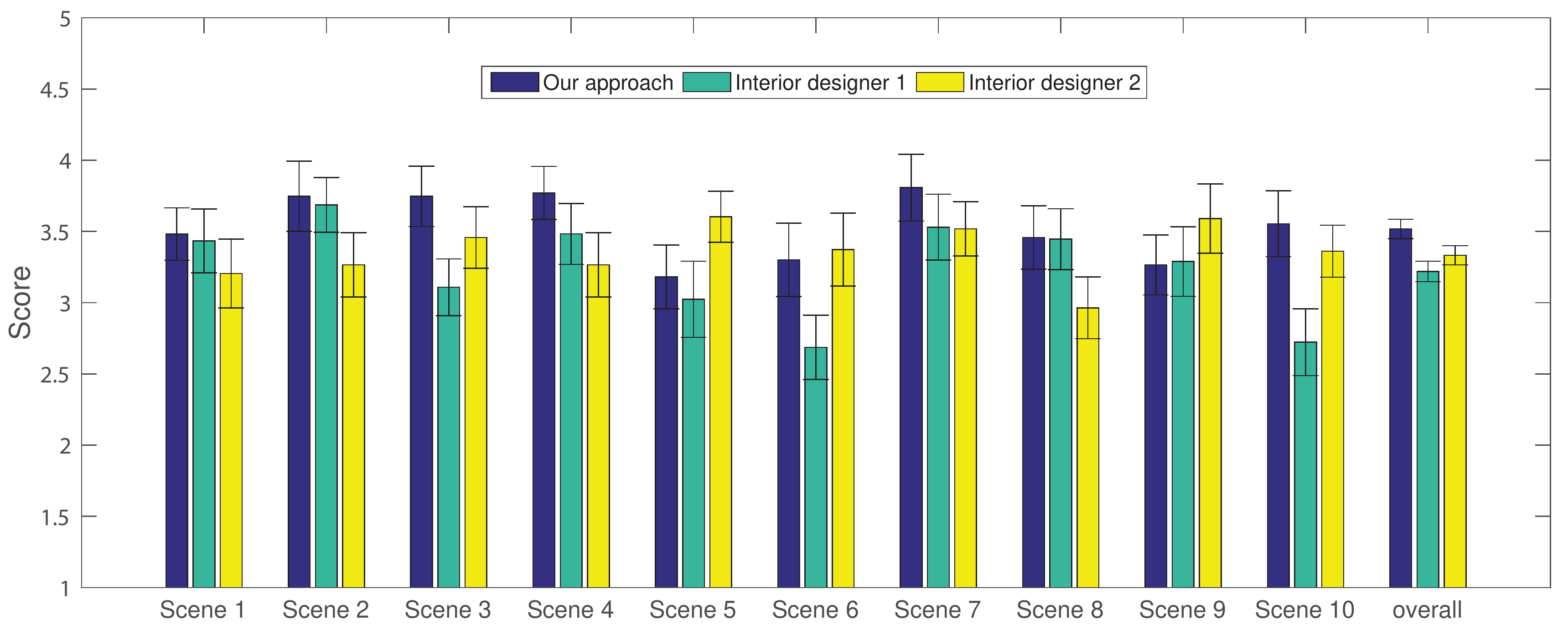}
  \caption{Average user ratings along with the 95\% confidence intervals of the results generated by our approach and those by 2 interior designers.}\label{UserStudy}
 \end{figure*}

\textbf{Study details.} There are totally ten scenes used in the study. The scenes, including the bedroom, dinning room, living room, meeting room, and office, were created using 3ds Max. We colorize each scene using our proposed approach with a specified target color theme as the input. We manually select 5 colors to make a visual-pleasing color theme as the input, but there are several methods which could help generating a harmonic color theme \cite{cohen2006color,o2011color}, which can be further incorporated into our system. The interior designers choose his/her favorite materials and textures by using this color theme as the reference as well. For each scene, the results by our approach and two designers were rendered with V-Ray 3.3 for 3ds Max under exactly the same viewpoint and illumination condition. The rendered scenes used for this study could be found in our supplemental material.

We have recruited 83 subjects for this task. Most subjects were the undergraduate or master students majored in Computer Science, but had a minimal graphics background. During the study, each subject saw all the ten groups of images. We ensure that all scenes are presented in a randomly permuted order, and that the image placement (left, middle, or right) is randomized. The subjects were asked to rate each image with a score between 1 (poor) and 5 (excellent), which indicates the degree of visual satisfaction over the results.


Figure \ref{UserStudy} shows the average rating along with the 95\% confidence interval on each group of the results. It has been shown that our score slightly outperforms the two interior designers on 7 groups. We further analyze the statistical data. For each group, if the score given by a subject is higher than the other two, we can safely assume that the subject prefers our result over the other two on this group. Overall, more than $55.64\%$ subjects prefer our results over those by the two designers.
As we talked with the designers we invited,
although our texture database contains 2650 texture samples, the designers still think that these textures are not enough which limit their choices to some extent.
They expect to see a much wider variety of textures which should be organized according to different design styles.
Besides, they also think the specified palettes limit their choices, but not a big problem.
However, this user study shows that most subjects can not distinguish the results produced by the machine automatically and those made by interior designers,
and they even prefer our results in most cases, which still demonstrate the effectiveness of our approach.

\begin{table}[htbp] \footnotesize
\centering
\tabcolsep11pt
\caption{Time costs of our approach and the interior designers.}
\begin{tabular}{|c|c|c|c|}
\hline
Scene No.    & Our approach & Designer 1 & Designer 2 \\ \hline
1            & 2.31s        & 27m48s              & 19m54s              \\ \hline
2            & 3.38s        & 31m16s              & 26m57s              \\ \hline
3            & 3.16s        & 23m37s              & 24m5s               \\ \hline
4            & 3.32s        & 25m7s               & 23m32s              \\ \hline
5            & 3.46s        & 27m57s              & 24m54s              \\ \hline
6            & 3.1s        & 23m35s              & 22m17s              \\ \hline
7            & 2.71s        & 26m24s              & 26m39s              \\ \hline
8            & 3.18s        & 29m59s              & 27m10s               \\ \hline
9            & 3.08s        & 24m6s               & 25m22s              \\ \hline
10           & 2.78s        & 28m12s              & 24m29s              \\ \hline
\end{tabular}
\label{UserStudyTable}
\end{table}

As shown in Table~\ref{UserStudyTable}, we also recorded the time used by each interior designer to process the input 3D scenes and select the materials for each furniture object, not including the time for rendering. On average, it takes at least 20 minutes for each scene. We talked with the invited designers after finishing the task. They said that even though they are experienced interior designers and are skilled in using 3ds Max, carefully selecting their favorite textures and assigning them to those furniture models with many geometric components are indeed a painful and time-consuming task. By contrast, the run time of our approach was approximately 3 seconds per 3D scene.

In conclusion, the user study shows that our approach could be a useful tool in interior design. It helps users relieve from the time-consuming task of furniture and scene decoration.


\section{Conclusions}
We have presented a novel approach that automatically generates color suggestions for 3D indoor scenes. Our approach is object-based in the sense that we have the capacity to colorize an independent furniture object according to furniture images. The core is to learn the image-guided model segmentation. In addition to recommending colorization schemes according to user's preferences, the object-based property of our approach also enables us to colorize the input scene by transferring colors from a user-provided indoor image, and furthermore, allows the user to refine the colorization scheme of the whole scene by interactively adjusting the colorization of a certain object. Our approach works in a data-driven manner. An image-model database with labeled hierarchical information and detailed image-model correspondences, along with a texture database, is contributed to the community.
Our results have been deemed by human observers to be perceptually close to the colorization results by human designers. People even cannot differentiate our automatically created results with those by interior designers.

Currently, we do not explicitly consider any high-level guidelines and aesthetic rules in interior and graphic design. Since existing professional photographs intrinsically follow these guidelines, working in a data-driven manner our approach essentially obeys them as well. Our experimental results and the pilot study validate our effectiveness. However, future work could extend our approach to explicitly integrate such criteria, which could be expressed as additional terms, into our objective function. Given a new indoor image as the reference, we have to label it first to make the 3D scene follow its colorization scheme. Though we have developed a prototype to do this, the task could be boring for novices. Scene understanding, and specifically scene parsing and object recognition, have achieved significant progress in recent years~\cite{russell2009associative,cheng2014imagespirit}. In the future, we plan to incorporate these into our prototype to assist user interaction.


%



\ifCLASSOPTIONcompsoc
  \section*{Acknowledgments}
\else
  \section*{Acknowledgment}
\fi

The authors would like to thank...

\ifCLASSOPTIONcaptionsoff
  \newpage
\fi



\bibliographystyle{IEEEtran}
\bibliography{template}

\end{document}